\documentclass[%
 aip,
 amsmath,amssymb,
 reprint,%
]{revtex4-1}

\usepackage{graphicx}
\usepackage{dcolumn}
\usepackage{bm}
\usepackage{xcolor}

\usepackage[utf8]{inputenc}
\usepackage[T1]{fontenc}
\usepackage{mathptmx}
\usepackage{etoolbox}
\usepackage{ulem}
\usepackage{tabularx}
\usepackage{booktabs}
\usepackage{float}

\makeatletter
\def\@email#1#2{%
 \endgroup
 \patchcmd{\titleblock@produce}
  {\frontmatter@RRAPformat}
  {\frontmatter@RRAPformat{\produce@RRAP{*#1\href{mailto:#2}{#2}}}\frontmatter@RRAPformat}
  {}{}
}%
\makeatother
\begin{document}

\preprint{AIP/123-QED}

\title{Quantifying Chirality in Helical Polymers via a Geometric Extension of the Kremer–Grest Model}
\author{Michael J. Grant}
 \affiliation{Department of Microsystems Engineering, Rochester Institute of Technology}
 
\author{Poornima Padmanabhan}%
 \email{poornima.padmanabhan@rit.edu.}
\affiliation{Department of Microsystems Engineering, Rochester
  Institute of Technology}
\affiliation{ 
Department of Chemical Engineering, Rochester Institute of Technology.
}%

\date{\today}

\begin{abstract}
Chirality in polymeric systems enables a wide range of emergent optical, mechanical, and transport phenomena, yet a unified framework that quantitatively connects molecular-scale geometry to chiral behavior remains lacking. Existing theoretical descriptions typically emphasize either continuum models, such as the helical wormlike chain (HWLC), which neglect intermolecular interactions, or mesophase-level theories, which obscure the role of molecular geometry. In this work, we introduce a comprehensive framework for quantifying chirality in helical polymers by extending the Kremer-Grest bead–spring model to explicitly map intrinsic curvature and torsion onto bond angle and dihedral potentials. We establish direct theoretical relationships between helical parameters such as pitch and radius, and connect them to a normalized, dimensionless chirality characteristic, $\chi$ that captures local geometric correlations absent from conventional HWLC descriptions. Furthermore, using molecular dynamics simulations, we systematically quantify the influence of excluded volume interactions and thermal fluctuations on helical geometry and chirality, dispelling the common assumption that monotonic increases in chirality are associated only with decreasing pitch. Finally, we present a coarse-graining procedure that facilitates a direct comparison between experimental helical polymers and the Kremer–Grest helical chain, demonstrating quantitative agreement across a diverse set of polymer classes. This unified geometric and particle-based description provides a predictive roadmap for selecting and engineering chiral Kremer–Grest models and offers a general platform for designing polymeric materials with controlled and tunable chirality.
\end{abstract}

\maketitle

\section{Introduction}
The geometry of helical polymers is inherently chiral, which often leads to a range of remarkable phenomena. Chiral polymers are well known for their light–matter interactions,\cite{Ennist2022,Zhao2019,Zhao2021,Zhong2023}  but their geometry also plays a critical role in hierarchical self‑assembly.\cite{Park2022,Hase2009,Wang2019,vanOosten2025} The helical conformation, in turn, enhances performance across diverse applications, including drug discovery, crystallization, asymmetric catalysis and polymerization, sensing, and the design of synthetic biomaterials.\cite{Miao2023,Ahmed2022,Inoue2023,Narmon2021,Reggelin2004,Chongsiriwatana2008} The diversity of polymer chemistries that give rise to helical conformations makes these systems challenging to study, underscoring the need to link helical geometry and the underlying polymer physics.\\

The use of Kremer-Grest (KG) model in molecular dynamics simulations has enabled us to study universal polymer behavior related to self-assembly and rheology. \cite{Rouse, PhysRevA.33.3628, Kremer1990, Kremer1992} Polymers are simply modeled by connecting hard-spheres with springs. These spheres interact through repulsive Lennard-Jones type interactions, with a bond length parametrized such that connected chains cannot pass through one another. For nearly four decades, it has been used to study the dynamics, rheology and self-assembly of an array of polymer-based materials.\cite{Duering1994, Murat1999, Halverson2012, Glaser2014, Seo2015, Svaneborg2020, Herschberg2021, Wang2024}\\

Yet, not all polymers are inherently flexible; some adopt semi-flexible or even rod-like conformations. Faller \textit{et al.} introduced angular potentials that enabled the KG model to be extended to study semi-flexible chains.\cite{Faller1999, Faller2000, Faller2001} This enhanced the utility of the model while maintaining its simplicity, and nearly twenty years after this addition, it was shown that this selective tuning of chain stiffness permits its use in modeling melts of a diverse variety of polymer chemistries. \cite{Everaers2020}\\

Progress in modeling helical polymers has taken several approaches. Yamakawa's helical wormlike chain model (HWLC) is based on describing the potential energy of a continuous helical curve, and offers a deep insight into the chain statistics of these chains.\cite{HWLC} A discrete version with bonds, curvature, and torsion was also developed, but due to a lack of excluded volume interactions, one cannot easily adapt it to study self-assembly. Orientational Self-Consistent Field Theory (oSCFT) has been highly successful in accounting for the role of chiral interactions in block polymer self-assembly, but lacks molecular level insights that govern polymer conformations. \cite{Zhao2012, Zhao2013, Grason2015} \\

To study helical polymers, the Kremer-Grest model (KG) has been augmented with dihedral angle potentials to drive the formation of a helix.\cite{Boehm2014, Glagolev2015, Buchanan2022} The inherent excluded volume interactions has made it possible to study the self-assembly of several helical polymers through molecular dynamics simulations.\cite{Glagolev2021,Grant2024,SpringsInStripes} \\

However, these studies have typically explored only a limited range of helical geometries with limited geometric characterization. At the same time, experimental work often emphasizes structural parameters such as pitch $h$ or the number of monomers per turn $m_t$. Together, these gaps point to an opportunity to better connect the geometric descriptions used in experiments with the representations of helicity employed in Kremer–Grest models of helical polymers.\\

\begin{figure}[ht]
    \centering
    \includegraphics[width=1\linewidth]{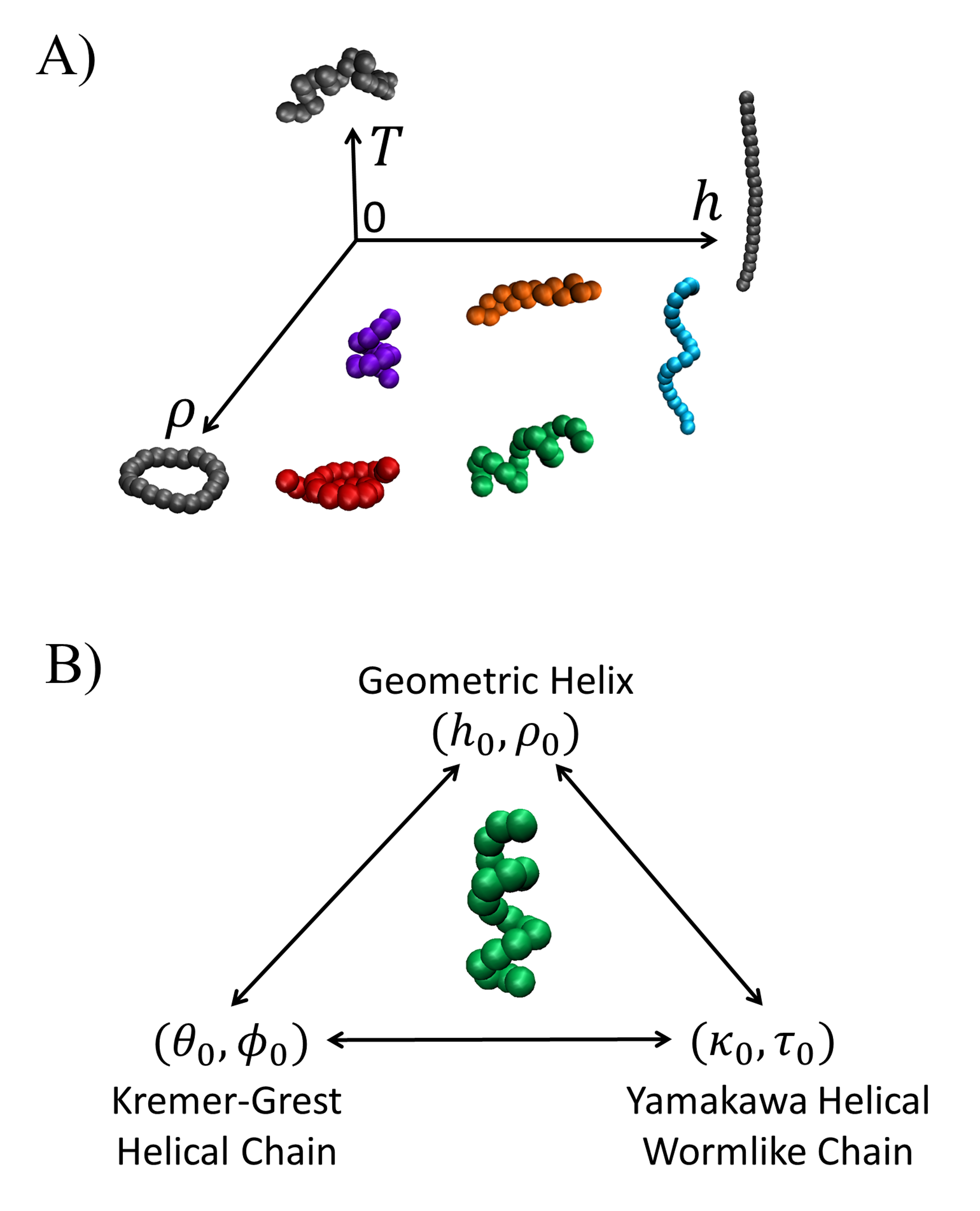}
    \caption{(A) The number of helical structures accessible is vast and can be parametrized by helical pitch $h$ and radius $\rho$. (B) This work establishes a framework that uses the tools of differential geometry ($\kappa_0$ and $\tau_0$) to develop a connection between the parametrization of an ideal helix and the Kremer-Grest helical chain, enabling a coarse-grained mapping between experimental polymers and bead-spring models.}
    \label{fig: introduction}
\end{figure}

Experimental work or geometric descriptions of helices often emphasizes structural parameters such as pitch $h$ or number of monomers per turn $m_t$ or helical radius $\rho$. Buchanan \textit{et al.} showed that the KG helical chain model can produce a vast array of helical conformations (Figure~\ref{fig: introduction}A) based on the choice of parametrization,\cite{Buchanan2022} but a detailed connection between the KG parameters and geometry is missing. In this manuscript, we reconcile several seemingly disparate models for helical polymers (Figure~\ref{fig: introduction}B). The HWLC by Yamakawa and coworkers describes a helical polymer by the curvature $\kappa_0$ and torsion $\tau_0$. The Kremer-Grest helical chains are parametrized by angular and dihedral setpoints, $\theta_0$ and $\phi_0$, respectively. Geometric descriptions of a helix and structural characterization in experiments are often reported in terms of helical pitch $h_0$ and radius $\rho_0$. To connect helical geometry to chirality, a chirality characteristic ($\chi_0$) is defined. The KG helical chain is characterized across its entire parameter space. Finally, we present a coarse-graining roadmap that facilitates a mapping between experimental helical polymers and the corresponding Kremer–Grest helical chain.

\section{Methods}
In this section, three distinct viewpoints of a helix and their interconnections are described in detail. In Section II.A., the continuous and discrete versions of Yamakawa's Helical Wormlike Chain Model in terms of curvature and torsion $(\kappa_0,\tau_0)$ are described. In Section II.B., the extension of the Kremer-Grest model to describe helical conformations parametrized by the angular and dihedral angles $(\theta_0,\phi_0)$ is presented. In Section II.C., a physically intuitive geometric description of a helix based on pitch and helical radius $(h_0,\rho_0)$ is examined. A normalized chirality characteristic $\chi$ is also defined to quantify the chirality of the shape.
\subsection{The Helical Wormlike Chain}
Yamakawa and coworkers developed the helical wormlike chain model (HWLC) to describe helical polymers.\cite{HWLC} In the model, the continuous chain is based on a parametric equation for the helix in terms of the curvature and torsion setpoints, $\kappa_0$ and $\tau_0$, respectively. To account for thermal fluctuations along the backbone, the potential energy  is defined as:
\begin{equation}
\label{eqn: helix}
    U_{helix}(\kappa,\tau) = \frac{\alpha}{2}\int_0^L (\kappa-\kappa_0)^2ds + \frac{\beta}{2}\int_0^L (\tau-\tau_0)^2ds,
\end{equation} 
where  $L$ is the contour length of the polymer, and $\kappa$ and $\tau$ denote the local curvature and torsion of the helix, respectively, constrained by the energetic parameters $\alpha$ and $\beta$.\\

Any point along a helical chain can be fully described with three orthonormal unit vectors; the tangent ($\mathbf{t})$, normal ($\mathbf{n}$) and binormal ($\mathbf{b}$). Together, these comprise a Frenet-Serret frame and are related to the curvature and torsion as: 
\begin{equation}
\label{eqn: kappa and kappa}
    \kappa=|\partial\mathbf{t(s)}/\partial s|\text{ and }\tau=\pm|\partial\mathbf{b}/\partial s|.
\end{equation}

The helical chain may be discretized by joining $N$ monomers separated by bonds, each of length $R_0$. Two successive monomers describe the tangent vector, three successive monomers are required to compute curvature $\kappa$, and four successive monomers are needed to compute torsion $\tau$ from the Frenet-Serret frames. The potential energy of a discrete helical wormlike chain is:
\begin{equation}
\label{eqn: discrete helix}
    U_{helix}(\kappa,\tau) = \frac{\alpha}{2}\sum_{i=3}^N (\kappa-\kappa_0)^2 + \frac{\beta}{2}\sum_{i=4}^N (\tau-\tau_0)^2
\end{equation} 

Yiu \textit{et al.} carried out a similar discretization and leveraged a transfer matrix method to determine the persistence length of the HWLC as a function of equilibrium curvature and torsion\cite{GreenHelix}. However, Equation~\ref{eqn: discrete helix} treats monomers as point particles and lacks non-bonded interactions which play a critical role in self-assembly and rheological properties of polymers. Next, we describe the Kremer-Grest helical model as described in Buchanan \textit{et al.}\cite{PhysRevA.33.3628, Buchanan2022}

\subsection{Kremer-Grest Model of a Helical Polymer}
Excluded volume interactions between monomers $i$ and $j$ are computed using a Weeks-Chandler-Andersen potential \cite{WCA}:
\begin{equation}
    U_{nb,ij}(r_{ij})=\begin{cases}4\epsilon\left[ \left( \frac{\sigma}{r_{ij}} \right) ^{12} - \left( \frac{\sigma}{r_{ij}} \right) ^{6} \right] + \epsilon, &r_{ij}< 2^{1/6}\sigma\\
    0, & r_{ij} \geq 2^{1/6}\sigma
    \end{cases}
\label{eqn: nonbonded}
\end{equation}
where $r_{ij}$ is the distance between monomers of index $i$ and $j$ and $\epsilon$ is the repulsion strength between the monomer types which is set to unity, $\epsilon=k_BT=1$. Connectivity of beads was enforced through a harmonic bond potential between successive monomers:

\begin{equation}
    U_{bond,ij}(r_{ij})=K_\text{har}(r_{ij}-R_0)^2,
\label{eqn: Bond}
\end{equation}
at an equilibrium bond length of $R_0=0.97\sigma$ and a force constant of $K_{har}=400k_BT$. \\

Helical conformations are obtained by implementing a bond angle potential and a dihedral potential. The bond angle potential is harmonic:
\begin{equation}
\label{anglePotential}
    U_{angle, ijk}(\theta_{ijk}) = K_\theta (\theta_{ijk} - \theta_0)^2,
\end{equation}
where $\theta_{ijk}$ represents the angle formed by three successive beads of indices $i$, $j$, $k$ and $\theta_0$ is the angular setpoint. The dihedral potential was implemented as a limited Fourier series expansion:

\begin{equation}
U_{dihed, ijkl}(\psi_{ijkl}) = K_\phi \left [ 1+\cos(\psi_{ijkl} - \psi_{0}) \right ]
\label{eqn: dihedralPotential}
\end{equation}
where $\psi_{ijkl}$ is the dihedral angle formed by four successive monomers with indices of $i$, $j$, $k$, $l$ and the overall energy is minimized at the dihedral setpoint, $|\psi_{ijkl}-180^\circ|=\phi_0$. Equation \ref{eqn: dihedralPotential} also dictates the handedness of the helix, where $\phi_0>0^\circ$ corresponds to right-handed helices and $\phi_0<0^\circ$ to left-handed helices. Furthermore, as $|\phi_0| > 90^\circ$ the polymer chain has a tendency to adopt a zig-zag conformation, rather than a compact helix.\\

We note that other functional forms of the bonded, angular, and dihedral potentials will also produce helical conformations as long as the force constants are sufficiently positive. As long as the linearization around the Kremer-Grest setpoints can be compared to the linearization from the HWLC, one can faithfully map the two models onto each other. The force constants $K_\theta$ and $K_\phi$ governing the helix inducing potentials are directly related to the parameters $\alpha$ and $\beta$ which control $\kappa$ and $\tau$, respectively. Within the HWLC framework, many of the relevant moments regarding chain statistics are derived under the constraint that $\alpha$=$\beta$.\cite{HWLC} To preserve this relationship in our simulations, we account for LAMMPS' implicit inclusion of $1/2$ in the definition of harmonic potentials, and set $2K_\theta = K_{\phi}=10k_BT$ to drive helix formation.\\

It is straightforward to show that $\theta_0$ and $\phi_0$ together constitute a measure of intrinsic $\kappa_0$ and $\tau_0$ respectively. We give explicit proofs of these relationships in Appendix A and summarize the interconversion between $(\kappa_0,\tau_0)\leftrightarrow (\theta_0,\phi_0)$ as:
\begin{equation}
\label{eqn: kappa}
    \kappa_0=\frac{\sqrt{2[1+cos(\theta_0)]}}{R_0},
\end{equation}
\begin{equation}
\label{eqn: tau}
    \tau_0 = \frac{sin(\theta_0)sin(\phi_0)}{\kappa_0^2R_0^3}.
\end{equation}
While these equations define the theoretical mapping between ($\theta_0,\phi_0$) and ($\kappa_0,\tau_0)$, in this work, particle positions are extracted from molecular dynamics trajectories and each Frenet-Serret frame from four successive monomers is used to compute $\kappa$ and $\tau$.\cite{SpringsInStripes} Due to (i) the excluded volume interactions and (ii) thermal fluctuations inherent to the simulations, the values deviate from the ideal predictions of equations \ref{eqn: kappa} and \ref{eqn: tau}. The reported values correspond to time-averaged measurements of all residues within the chain.

\subsection{Calculation of Geometric Properties}
The theoretical mapping between pitch and helical radius and curvature and torsion, $(h_0,\rho_0)\leftrightarrow (\kappa_0,\tau_0)$, is given as:
\begin{equation}
\label{eqn: pitch}
    h_0 = \frac{2\pi\tau_0}{\kappa_0^2 +\tau_0^2},
\end{equation} 
and the helical radius ($\rho_0$) is:
\begin{equation}
\label{eqn: rho}
    \rho_0 = \frac{\kappa_0}{\kappa_0^2 +\tau_0^2}.
\end{equation}
In simulations, the pitch and radius are computed from the instantaneous $\kappa$ and $\tau$ obtained from particle trajectories and the time-averaged value across all residues is reported.\\

To connect the helical geometry to the quantification of chirality, we implement a normalized chirality characteristic ($\chi\in[-1,1]$):\cite{Abraham}
\begin{equation}
\label{eqn: chirality characteristic}
    \chi = \frac{1}{M}\sum_{i=3}^{M}\frac{(\mathbf{t}_{i-2}\times \mathbf{t}_{i-1})\cdot \mathbf{t}_i}{|\mathbf{t}_{i-2}||\mathbf{t}_{i-1}||\mathbf{t}_{i}|}    
\end{equation}
where $M$ is the count of all bond vectors in the chain ($M=N-1$). Chirality vanishes either when the bonds are collinear (limiting $\rho$ or $\kappa\rightarrow0$) or coplanar (limiting $\tau$ or $h\rightarrow 0$). For an ideal helix, the numerator simply becomes $\tau_0\kappa_0^2$ (Equation~\ref{eqn:tauDerivation} in Appendix A) and the denominator is proportional to $R_0^3$, allowing the equilibrium $\chi_0$ to be written as:
\begin{equation}
    \chi_0 =\frac{1}{M}\sum_{i=3}^{M}\tau_{0,i}\kappa_{0,i}^2=\frac{1}{MR_0^3}\sum_{i=3}^{M}sin(\theta_0)sin(\phi_0).
\label{eqn: chi tau kappa}
\end{equation}
Here, Equation \ref{eqn: chi tau kappa} corresponds to a theoretical expression, while the reported results are obtained using \ref{eqn: chirality characteristic}, computed directly from extracted particle positions. The reported chirality characteristic corresponds to time-averaged measurements. \\

In Figure~\ref{fig: introduction}B, Equations 8 and 9 interrelate HWLC parameters to the KG helical chain and Equations 10 and 11 interrelate geometric descriptors to HWLC, thereby closing the loop for ideal chains. In addition, Equation 13 measures a normalized chirality characteristic. \\

To demonstrate the efficacy of our proposed mapping, we compare the results from molecular dynamics simulations of the Kremer-Grest Helical chain to the theoretical predictions computed in Equations \ref{eqn: kappa}-\ref{eqn: chi tau kappa}. We explore a suite of single polymer chain simulations where we systematically vary the equilibrium bond angle  $\theta_0\in[60^\circ,170^\circ]$ in increments of $10^\circ$, and the dihedral angle setpoint $\phi_0 \in[0^\circ,90^\circ]$ also in increments of $10^\circ$. For $\phi_0\in[-90^\circ,0^\circ]$ (data not shown), one obtains the opposite handedness. We then show how fluctuations and excluded volume effects influence polymer conformation and use these insights to demonstrate how this model can be used to elucidate the chiral behavior of real polymer chains.

\subsection{Simulation Details}
All simulations were conducted within the framework of the Large-scale Atomic/Molecular Massively Parallel Simulator (LAMMPS) using resources offered by Reasearch Computing at RIT \cite{LAMMPS2022, RITCOMPUTING}. Dimensionless units were invoked by setting the reference mass ($m$), length ($\sigma$) and energy ($k_BT$) to unity.\\

Simulations are run in the NVT ensemble with triply periodic boundary conditions and a Nos\'{e}-Hoover thermostat maintaining a temperature of $T^*=1 k_BT$. A single polymer of chain length $N=100$ is initialized as a random walk in the cell with a density of $\rho^*=0.001\sigma^{-3}$. A soft potential was used  to break initial overlaps between monomers for $57.5\tau$ before switching to the Weeks-Chandler-Andersen potential. The random coil is equilibrated for $500\tau$ before activating the bond angle and torsion potentials, where the helical polymer chain is further equilibrated for an additional $5,000\tau$. Finally, a production run of $5,000\tau$ is collected and the resulting trajectory analyzed.

\section{The Kremer-Grest Helical Chain}
\subsection{Theoretical Results}
\begin{figure}[ht]
    \centering
    \includegraphics[width=1\linewidth]{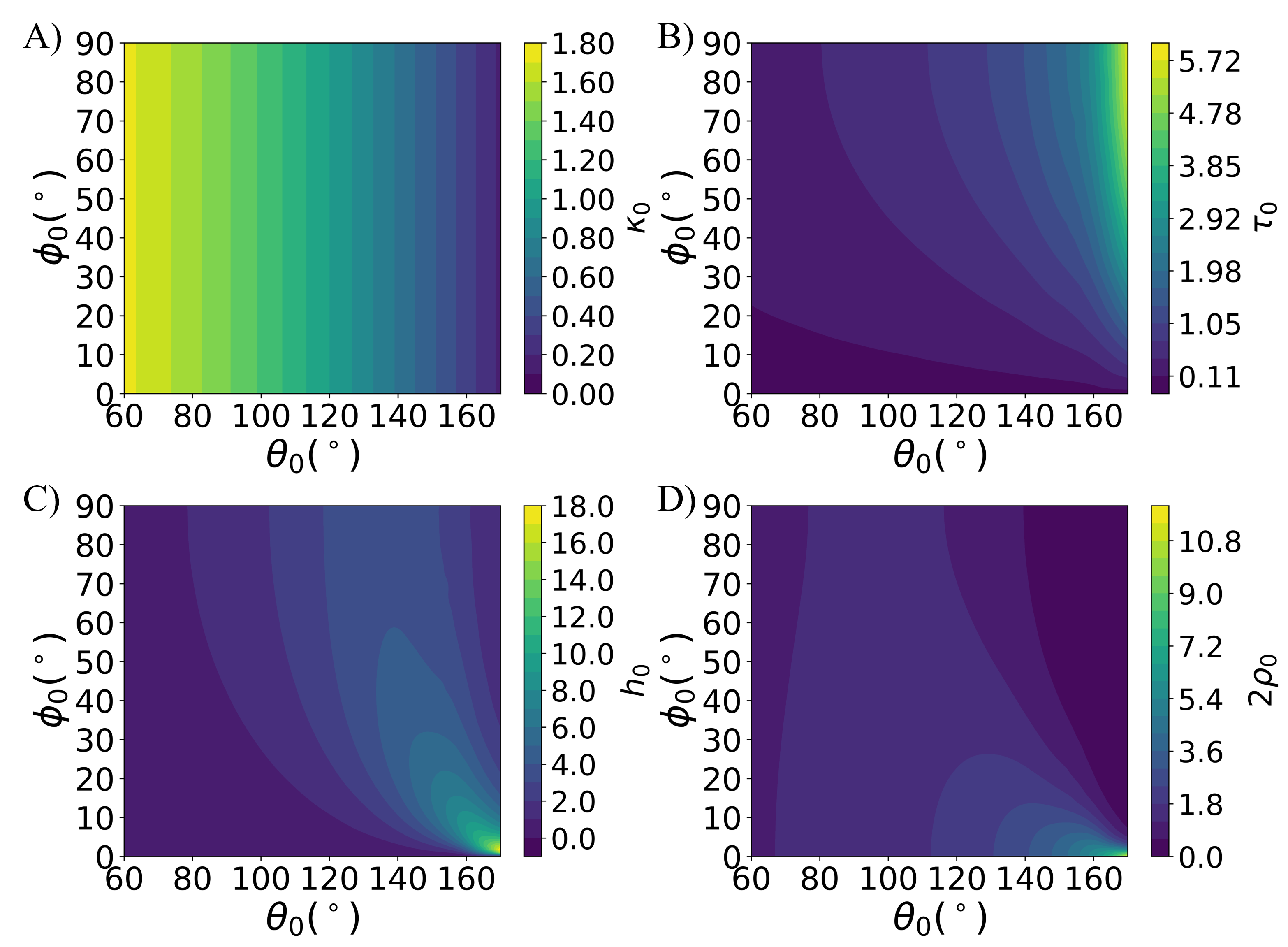}
    \caption{Theoretical values computed for (A) $\kappa_0$, (B) $\tau_0$, (C) $h_0$, (D) $2\rho_0$. Bond angles greater than $170^\circ$ result in undefined values of torsion due to high prevalence of collinearity.}
    \label{fig:Analytical kappa tau height and rho}
\end{figure}

We first start by examining $\kappa_0$, $\tau_0$, $h_0$ and $2\rho_0$ obtained from Equations \ref{eqn: kappa}-\ref{eqn: rho}. In Figure \ref{fig:Analytical kappa tau height and rho}A, curvature $\kappa_0$ increases monotonically as a function of $\theta_0$ with no dependence on $\phi_0$, precisely agreeing with Equation~\ref{eqn: kappa}. Calculations for torsion $\tau_0$ shown in Figure \ref{fig:Analytical kappa tau height and rho}B demonstrate that $\tau_0$ increases as a function of $\theta_0$ and maximizes when $\phi=90^\circ$, as expected from Equation~\ref{eqn: tau}. Pitch $h_0$ (Figure \ref{fig:Analytical kappa tau height and rho}C) follows similar behavior as $\tau_0$ until roughly $\theta_0 \approx 140^\circ$, above which it decreases for high $\phi_0$ and decreases for low $\phi_0$. This is explainable from Equation~\ref{eqn: pitch} where diminishing curvature at high $\theta_0$ leads to a $2\pi/\tau_0$ dependence. Similarly, helical diameter $2\rho_0$ (Figure~\ref{fig:Analytical kappa tau height and rho}D) increases with $\theta_0$ continually for low $\phi_0$, but for high $\phi_0$, it increases with $\theta_0$ and later decreases. This behavior follows Equation~\ref{eqn: rho} where in the low curvature regime, it scales as $\kappa_0$, but in the low torsion regime, it scales as $2/\kappa_0$.

\begin{figure}[ht]
    \centering
    \includegraphics[width=1\linewidth]{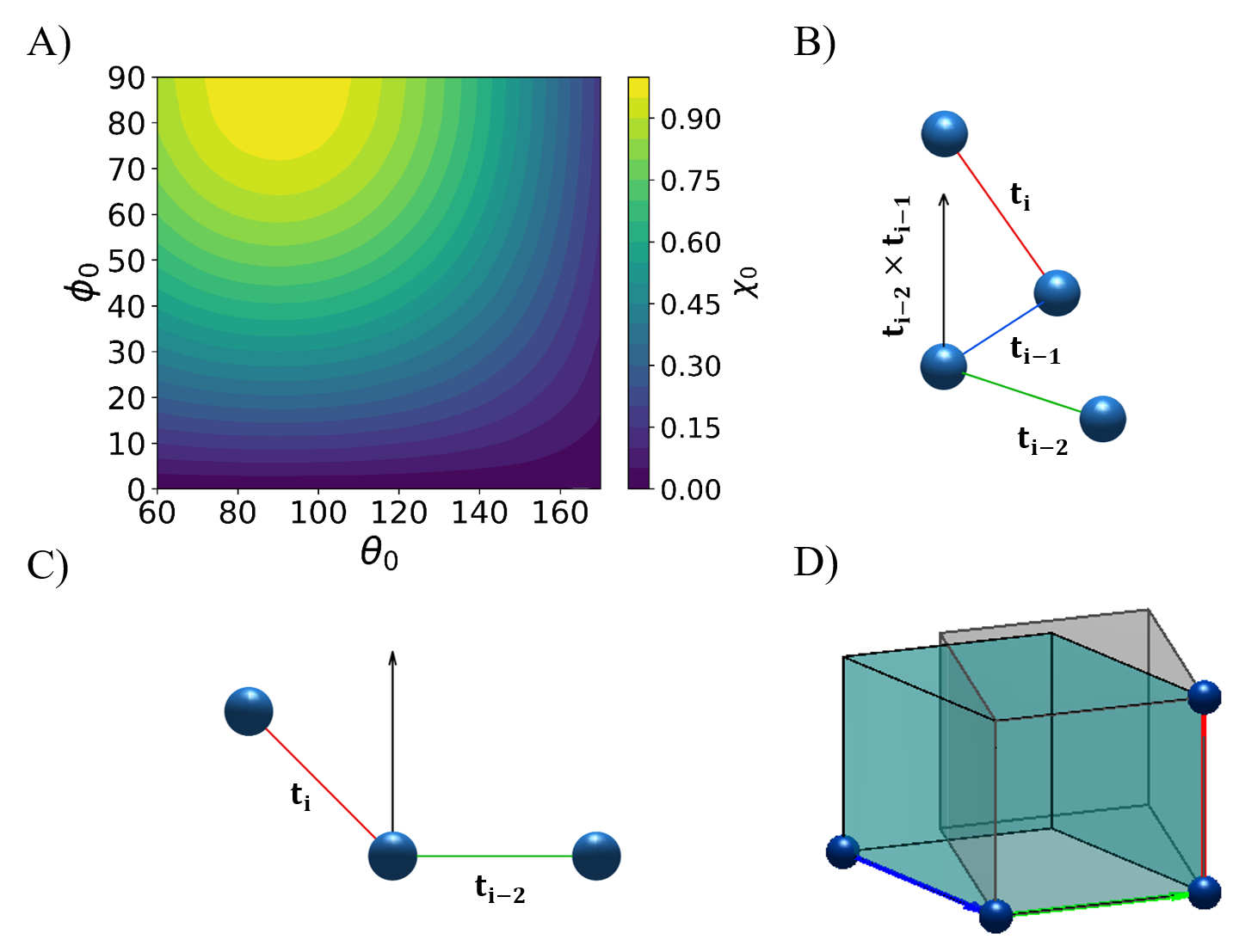}
    \caption{(A) Theoretical chirality characteristic ($\chi_0$) computed from Equation \ref{eqn: chi tau kappa}. (B) A representative helical residue similar to what would be extracted from molecular simulations. (C) The helical residue when projected along the $\mathbf{t}_{i-1}$ bond vector. (D) Perturbations to a parallelepiped volume when bonds in a helical residue are not mutually orthogonal.}
    \label{fig:analytical chirality}
\end{figure}

Next, we focus on the chirality characteristic ($\chi_0$), which measures the volume of the parallelepiped formed by successive bond vectors. It depends substantially on both curvature and torsion from Equation~\ref{eqn: chirality characteristic}, and in the $(\theta_0,\phi_0)$ plane (Figure~\ref{fig:analytical chirality}A), it follows a simple polar sine relation with a maximum at $(\theta_0=90^\circ,\phi_0=90^\circ)$ that radially decreases outward. To visualize this behavior, a representative set of vectors is shown in Figure~\ref{fig:analytical chirality}B, and the cross product $\mathbf{t}_{i-2}\times \mathbf{t}_{i-1}$ is related to the curvature or local bending. When the bonds are orthogonal, the cross product is maximized. As the bond angle deviates from $\theta=90^\circ$, the area and curvature decrease, vanishing entirely under collinearity. The same vectors are projected along the $\mathbf{t}_{i-1}$ vector in Figure~\ref{fig:analytical chirality}C to provide insight into the dihedral angle. At $\phi=90^\circ$, the relevant planes are perpendicular, maximizing local twist. At $\phi=0^\circ$, the relevant planes are coplanar, thereby causing chirality to vanish. Finally, the chirality characteristic can be visualized as the volume of the parallelepiped in Figure~\ref{fig:analytical chirality}D. An orthonormal set would result in a cube representing maximal chirality, while deviations from orthogonality reduce the volume (Figure 2D), and consequently reduce chirality.\\

In fact, comparing the location of maximum chirality to the trends in helical pitch and radius in Figure~\ref{fig:Analytical kappa tau height and rho}, it becomes apparent that to maximize chirality, moderate values of pitch and helical diameter are necessary. This insight departs from other views in the literature that simplify the measure of chirality as inverse pitch.\cite{Grason2015}

\subsection{Excluded Volume Effects}

\begin{figure*}[t]
    \centering
    \includegraphics[width=1.0\linewidth]{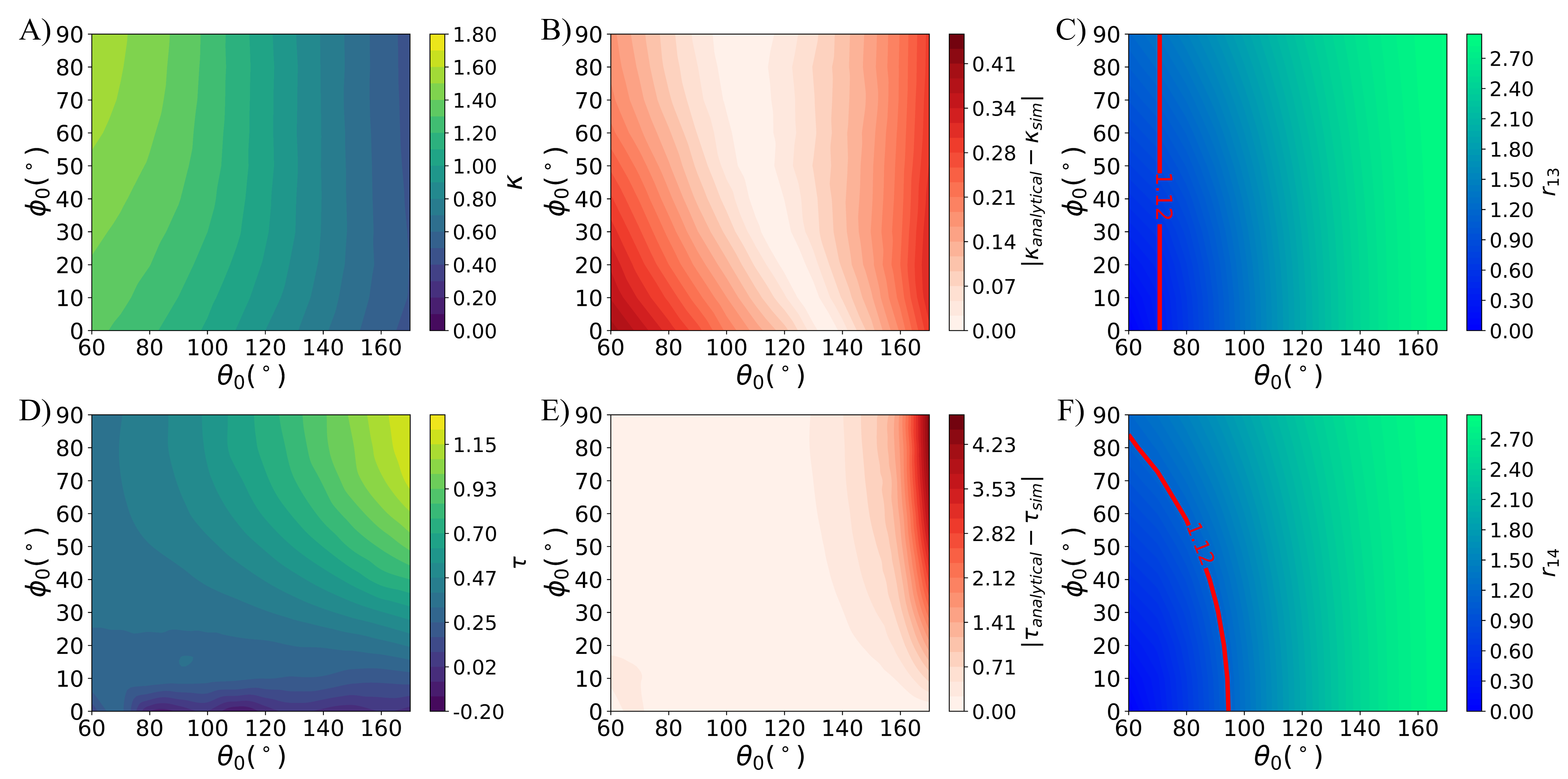}
    \caption{(A) Local curvature $\kappa$ derived from molecular dynamics simulations trajectories. (B) Deviation of local $\kappa$ (simulation) from theoretical $\kappa_0$. (C) Distance between 1-3 monomer pairs, with repulsive cutoff overlaid. (D) Local torsion $\tau$ derived from molecular dynamics simulation trajectories. (E) Deviation of local $\tau$ (simulation) from theoretical $\tau_0$. (F) Distance between 1-4 monomer pairs, with repulsive cutoff overlaid.} 
    \label{fig:kappa tau error}
\end{figure*}


The Kremer-Grest helical model exhibits appropriate excluded volume interactions while conforming to a helix, which could lead to departures from the theoretical equations for the HWLC and geometric descriptors. In this section, we quantify the chain properties in simulation and compare them to the theoretical predictions. Interestingly, the reasons for departure from the theoretical model arise from several different factors.\\

Figures \ref{fig:kappa tau error}A and D show the values of $\kappa$ and $\tau$ extracted from molecular dynamics simulations respectively, by computing the local Frenet-Serret frame for each particle along the chain. Figures~\ref{fig:kappa tau error}B and~\ref{fig:kappa tau error}E show the deviation from theoretical predictions. For curvature $\kappa$, the magnitude in error is small, and the general trend remains similar to the theoretical prediction in Figure~\ref{fig:Analytical kappa tau height and rho}A, except for sharp deviations observed in the region of small $\theta_0$ and $\phi_0$ or high $\theta_0$. A contribution to this error stems from the 1-3 particle pair interactions. If the angle between successive bonds is too small, the repulsive non-bonded forces push the two monomers away from one another and prevent the residue from reaching the setpoint $\theta_0$ (and equivalently $\kappa_0$). Yet, the impact of 1-3 excluded volume interactions only occurs at $\theta_0\lesssim71^\circ$ (Figure 4C, see Appendix C for derivation), indicating that the error at larger bond angles must arise from other particle pair interactions. Specifically, the distance between particles 1 and 4 has a nonlinear dependence on both $\theta_0$ and $\phi_0$ (Figure 4F; the derivation can be found in Appendix D). The region in which the distance between both 1-3 and 1-4 particle pair interactions is less than the non-bonded cutoff corresponds to the deviations of the Kremer-Grest helical chain from the ideal conformations. \\

The deviation of torsion $\tau$ from theoretical values is larger in magnitude, and occurs predominantly at large $\theta_0$, and to a smaller extent when $\theta_0\lesssim 71^\circ$ (Figure \ref{fig:kappa tau error}E). This behavior as $\theta_0 \rightarrow 180^\circ$ arises from the high energetic penalties associated with collinearity and zero curvature of successive bond angles: if the polymer were to adopt this conformation, torsion becomes undefined and the simulation becomes unstable. To avoid this highly unfavorable state, the polymer adopts a configuration with higher curvature, which ultimately reduces the pitch and radius. While this is also observed in the error for $\kappa$, it is much more pronounced for $\tau$ due to the inverse squared dependence on curvature.\\

Beyond 1-3 and 1-4 interactions, the helical chain shape itself can lead to excluded volume interactions between monomers along the backbone. At small dihedral angles the helical residue adopts a coplanar conformation, resulting in a diminished pitch. Bond angles less than $\sim90^\circ$, coupled with the minimal rise of successive monomers, again results in a repulsion between non-bonded monomers preventing the residue from adopting its ideal conformation.\\

These errors arising from non-bonded interactions propagate to all derived helical properties. To demonstrate this, we compare three models, one with minimal error, one with high error in $\kappa$, and finally, one with significant error in both $\kappa$~and~$\tau$.\\

First, we examine the model with setpoints at $\theta_0 = 90^\circ$ and $\phi_0 = 80^\circ$, corresponding to a regime of minimal error. In this case, small deviations in $\kappa$ and $\tau$ have a negligible effect on the resulting polymer pitch and radius. By contrast, when a model exhibiting significant errors in both $\kappa$ and $\tau$ is considered ($\theta_0 = 170^\circ$ and $\phi_0 = 90^\circ$), the error in pitch increases dramatically, whereas the error in radius remains relatively modest. This pronounced pitch error originates from the theoretical model’s substantial underestimation of the pitch, a direct consequence of neglecting excluded-volume effects between monomers along the backbone. \\

Finally, we consider a model characterized by a notable error in $\kappa$ but minimal error in $\tau$. In this case, the error in pitch is less severe than in the previous model, but the error in radius becomes significantly more pronounced, a result of the theoretical prediction underestimating the true radius. \\

\begin{table}[ht]
  \begin{tabular}{c c c c c c}
    \hline
    $\theta_0$ & $\phi_0$ & $\kappa_{error}$ & $\tau_{error}$ & $h_{error}$ & $\rho_{error}$\\
    \hline
    $90^\circ$ & $80^\circ$ &  0.05 & 0.00 & -0.09 & -0.02 \\
    $170^\circ$ & $90^\circ$ &  -0.29 & 4.50 & -3.43 & -0.28\\
    $70^\circ$ & $10^\circ$ &  0.32 & -0.24 & -0.88 & -0.11 

    \\
    \hline
  \end{tabular}
    \caption{Simulation $\theta_0$ and $\phi_0$ setpoints and associated error in $\kappa$, $\tau$, $h$, and $\rho$ to the theoretical model.}
  \label{tbl:errors}
\end{table}

Excluded volume interactions induce systematic deviations from ideal helical conformations, yet the effects of thermal fluctuations have been ignored so far. These constitute a dominant factor in modulating the polymer geometry by continuously displacing residues from their equilibrium states. Thus, understanding how this affects helical geometry is pivotal in developing a model capable of simulating systems with  controlled chirality.

\subsection{Thermal Fluctuations}
In the HWLC, $\alpha$ and $\beta$ (Equation \ref{eqn: helix}) regulate the strength of helical driving forces for $\kappa$ and $\tau$, respectively. In the limit of large thermal fluctuations, where $\alpha\text{ and }\beta \ll k_BT$ these force constants are effectively negligible, leading to conformations that resemble a random coil with a comparatively short persistence length. By contrast, when $\alpha\text{ and }\beta > k_BT$, the influence of $\alpha$ and $\beta$ becomes pronounced, resulting in a semi-flexible helical conformation with a substantially increased persistence length and chain anisotropy. \\

Similarly, in the Kremer-Grest chiral model, these driving forces are governed by $K_\theta$ and $K_\phi$ (Equations \ref{anglePotential} and \ref{eqn: dihedralPotential}). In the limit of $2K_\theta=K_\phi\equiv K=0k_BT$, the original Kremer-Grest model is recovered and the resulting bond and dihedral angles adopt the conformations of a self-avoiding random coil. As the force constants increase, the helical chain shape emerges, and the persistence length also increases. In a helix, the winding nature of the coil necessitates significantly large values of degree of polymerization to reliably measure persistence length and this is not the focus of this work. Instead, we examine the role of thermal fluctuations in the transition from random coil to the helical conformation, with particular attention to those in Table~\ref{tbl:errors} that exhibit varying degrees of error between measured and theoretical $\kappa$ and $\tau$.\\

\begin{figure*}[ht]
    \centering
    \includegraphics[width=0.7\linewidth]{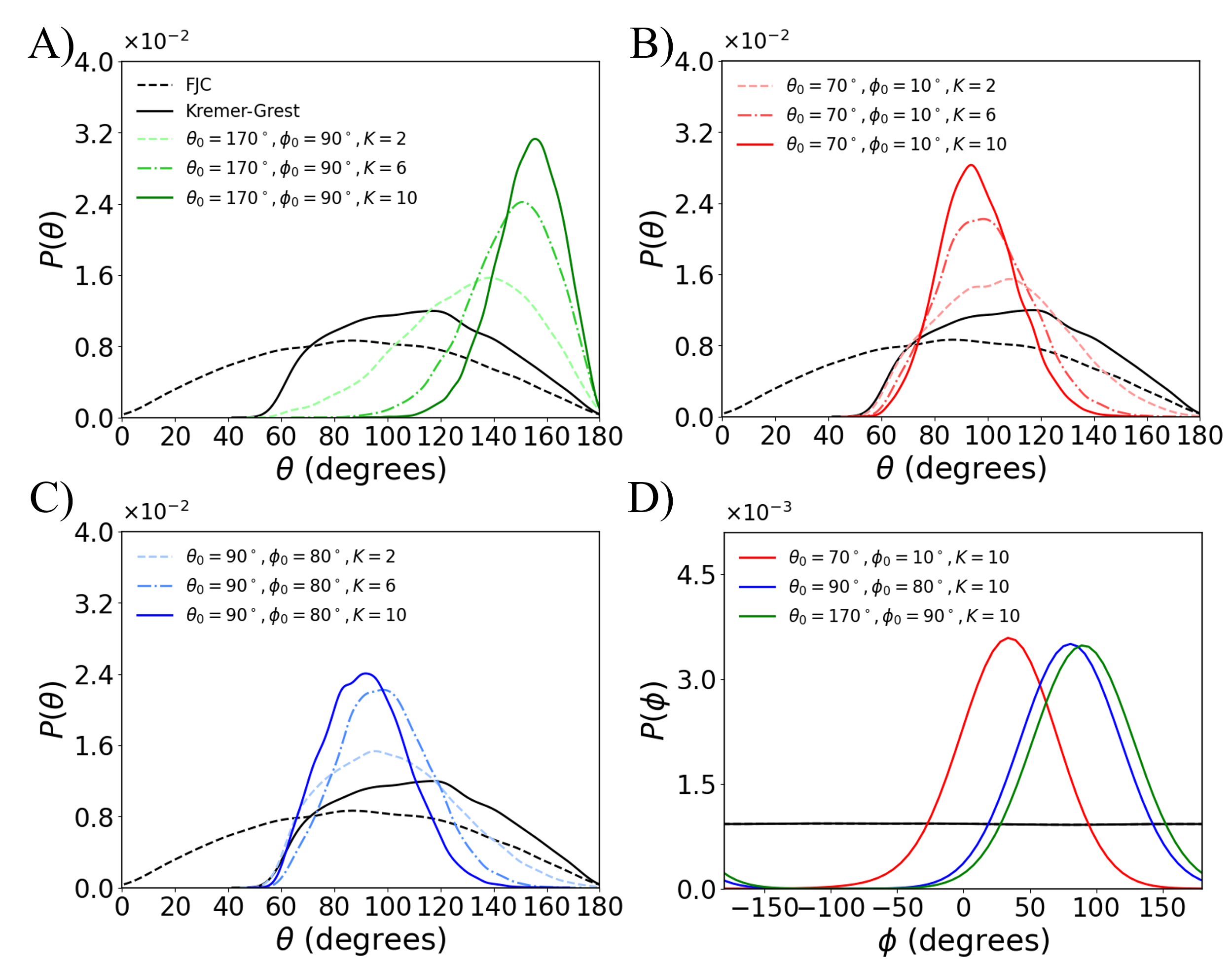}
    \caption{(A) Bond angle ($\theta$) distributions and (B) Dihedral angle ($\phi$) distributions for a freely jointed chain, standard Kremer-Grest model, and  select chiral models for $2K_\theta = K_\phi\equiv K=10k_BT$. (C) Impact of increasing $K/k_BT$ on $\kappa$. (D) Impact of increasing $K/k_BT$ on $\tau$.}
    \label{fig:thermal fluc}
\end{figure*}
To begin with, the bond and dihedral angle distributions of a freely-jointed chain (FJC), i.e. the only potentials governing the dynamics are the bonded potentials, are plotted in Figure~\ref{fig:thermal fluc}. Under these minimal constraints, conformational entropy dominates the system with distribution centered around $\theta=90^\circ$ (Figure~\ref{fig:thermal fluc}A--C) and a uniform distribution of $\phi$ (Figure~\ref{fig:thermal fluc}D). As discussed in previous work, if a reference bond connecting the center of a unit sphere and the pole forms a bond angle of $90^\circ$ with another bond connecting the center of a unit sphere, the second bond traverses the equator, maximizing the available conformations and thereby maximizing entropy.\cite{Grant2024} Next we consider the standard Kremer-Grest model with repulsive non-bonded interactions. This leads to relatively no change in the distribution of $\phi$ while $\theta$ is unable to adopt values below $60^\circ$ due to large 1-3 excluded volume interactions. \\

Upon increasing $K$ from 0 to 10$k_BT$, one expects that the angular distributions gradually shift from the Kremer-Grest values to a peaked distribution around the setpoint as the trade-off between entropic driving forces and enthalpic penalties shifts. For $\theta_0=170^\circ,\phi_0=90^\circ$ (Figure~\ref{fig:thermal fluc}A), the distribution shifts toward the right, but peaks well below the setpoint at $\theta_0\approx 160^\circ$, even though thermal fluctuations are significantly suppressed at $K=10k_BT$. As explained in Section III.B, no excluded volume interactions between 1-3 or 1-4 monomer pairs play a role. Rather, the trade-off that the molecules exhibit correspond to avoiding lower entropy associated with a nearly collinear chain at the setpoint $\theta_0=170^\circ$ in favor of larger enthalpic penalties for deviating from the angular setpoints, even at $K=10k_BT$.\\

In Figure~\ref{fig:thermal fluc}B, the angular distribution for $\theta_0=70^\circ,\phi_0=10^\circ$ shifts towards the left (and toward the setpoint), but the peak does not reach the setpoint. Here, the 1-3 and 1-4 repulsions play a significant role in dictating behavior rather than the thermal fluctuations. In Figure~\ref{fig:thermal fluc}C, the distribution for $\theta_0=90^\circ,\phi_0=80^\circ$ peaks around the setpoint even for $K\approx 6k_BT$. Here, the enthalpic penalties are traded off at the expense of loss of some entropy. \\

Given its ability to account for excluded volume effects and thermal fluctuations, the Kremer-Grest chiral model is sufficient for capturing the conformations of helical polymers governed by the HWLC. Therefore, it is critical to evaluate how such sources of error influence the chirality characteristic, $\chi$.

\subsection{Chirality Characteristic}
It is necessary to tie the helical geometry to a measure of chirality because of the impact on chiral mesoscale properties. Other studies in the literature have discussed a relationship between chirality and geometry. Within the framework of oSCFT, the strength of chirality is defined in terms of the inverse cholesteric pitch of the orientational field governing mesophase formation.\cite{Zhao2013} Although the relevant length scale is far larger than bead-spring models, it is worth noting that a dependence on the chain shape geometry is missing. In contrast, Yamakawa and others emphasized the intrinsic helical geometry of individual chains, alluding to maximizing `helical nature' when the ratio $\dfrac{\pi\tau_0}{\kappa_0}=\dfrac{h_0}{2\rho_0}\approx 1$, but did not explicitly quantify chirality.\cite{HWLC} Here, we measure the normalized chirality characteristic $\chi_0$ for the Kremer-Grest helical model. The chirality of the entire chain scales with the degree of polymerization and is given by $\chi_0N$, which would be related to experimental or DFT-based predictions of circular dichroism response of these materials.\cite{Rebholz2024} \\

\begin{figure}[ht]
    \centering
    \includegraphics[width=1.0\linewidth]{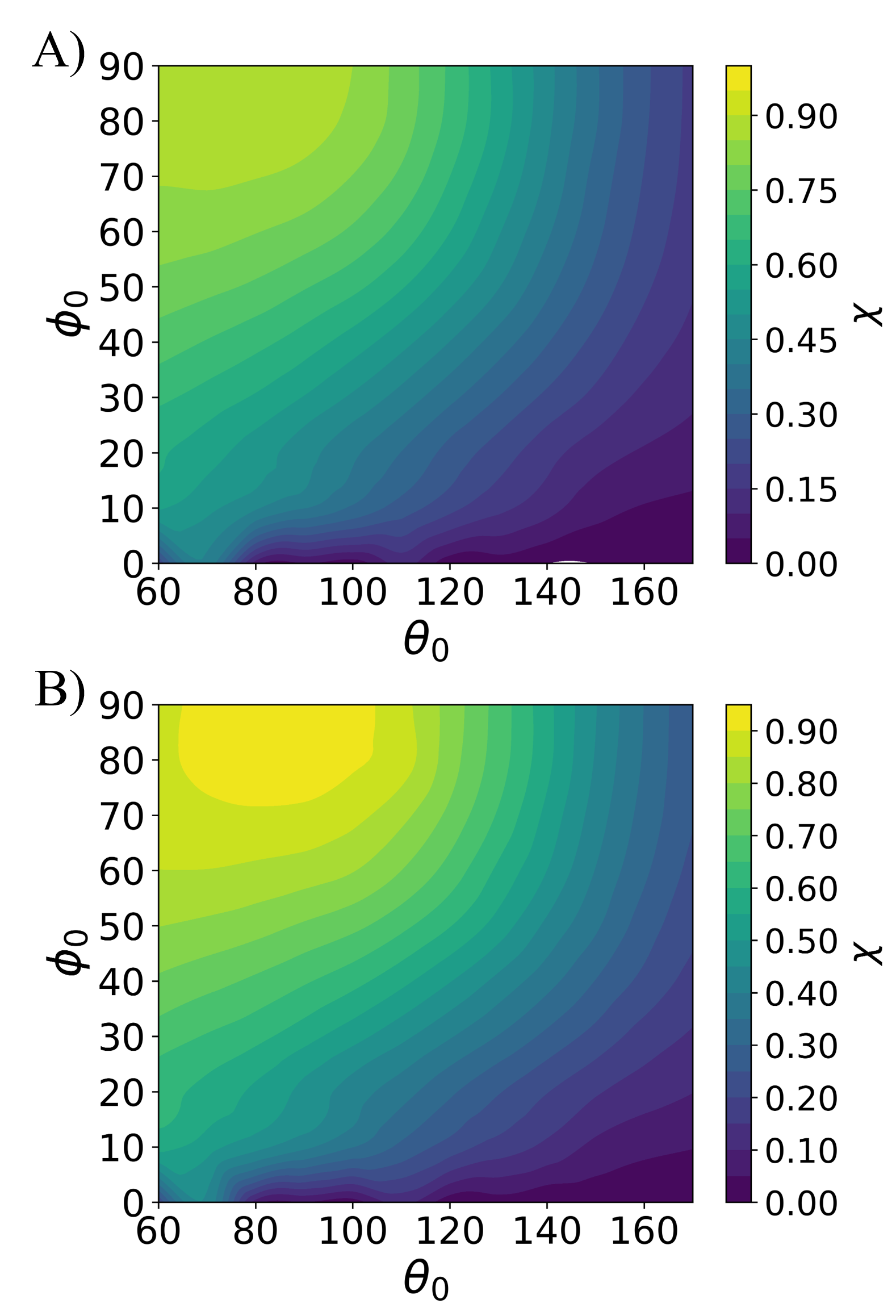}
    \caption{(A) $\chi$ computed using Equation \ref{eqn: chirality characteristic} and particle trajectories obtained in molecular dynamics simulations showing good agreement with our analytical model. (B) $\chi=\tau\kappa^2$ computed from the same trajectories.}
    \label{fig:simulated chirality}
\end{figure}

In Figure \ref{fig:simulated chirality}A, we use Equation \ref{eqn: chirality characteristic} to measure and plot the $\chi$ extracted from our molecular dynamics trajectories, finding good agreement with the analytical predictions (Figure~\ref{fig:analytical chirality}A). To relate to geometric properties, we invert Equations~\ref{eqn: pitch} and~\ref{eqn: rho} as:
\begin{equation}
\label{eqn: parametric kappa and tau}
    \kappa=\frac{\rho}{\rho^2+(h/2\pi)^2}, \tau=\frac{(h/2\pi)}{\rho^2+(h/2\pi)^2},
\end{equation} 
and use Equation 12 relating $\chi$ and $(\kappa,\tau)$ to get:
\begin{equation}
\label{eqn: expanded}
    \chi=\dfrac{1}{M}\sum_{i=3}^M\tau\kappa^2 = \dfrac{1}{M}\sum_{i=3}^M\frac{h\rho^2}{2\pi[\rho^2+(h/2\pi)^2]^3}.
\end{equation}

Figure~\ref{fig:simulated chirality}B plots the chirality characteristic using Equation~\ref{eqn: expanded} and is nearly identical to Figure~\ref{fig:simulated chirality}. Equation~\ref{eqn: expanded} provides a more nuanced relationship between $\chi\leftrightarrow(h,\rho)$ compared to other studies in the literature. Furthermore, in the limit of either $h$ or $\rho$ approaching $0$, the relationship shows that chirality is destroyed when the geometry is either a planar ring ($\tau=0=h$) or rigid rod ($\kappa=0=\rho$).\\

Equation~\ref{eqn: expanded} indicates that the chirality characteristic is inversely related to pitch as long as the radius does not collapse to zero, but perhaps more strongly than considered at the mesoscale using oSCFT. From Equation~\ref{eqn: expanded}, it turns out that maximum $\chi_0$ roughly occurs for $h_0/2\rho_0\approx 1.5$, somewhat agreeing with Yamakawa's prediction of maximal `helical nature'. Thus, the chirality characteristic measured in this study appears to be quite insensitive to errors in $\kappa$ and $\tau$ and agrees with geometric ideas in the literature.

\subsection{Geometry of Finite Chains}
Thus far, we have established two intrinsic length scales relevant to a helix: pitch and diameter. Although the discretization of the bead $R_0$ also plays a role, in this study, it has been held constant throughout. Pitch and diameter naturally lead to a dimensionless aspect ratio ($\Lambda_{turn}$):
\begin{equation}
    \Lambda_{turn} = \frac{h}{2\rho},
\label{eqn: Helical Turn Aspect Ratio}
\end{equation}
which characterizes the anisotropy of a helical turn. Coupled with the semi-flexible nature of helical chains, one can expect liquid-crystalline behavior.\cite{Onsager} Since the degree of polymerization $N$ can be a design parameter, the aspect ratio of the entire molecule $\Lambda_{helix}$ becomes an important chain characteristic. While $\Lambda_{helix}$ will scale with the degree of polymerization in a helix, the scale factor is nuanced due to helical winding. Specifically, the length is governed not only by $N$, but also by the number of monomers per turn, $m_t$, computed as the ratio of the arc length of a full helical turn ($s_{turn}$) to the arc length between two successive monomers ($s_{mon}$):
\begin{equation}
    \label{eqn: monomers per turn}
    m_t =\frac{s_{turn}}{s_{mon}}=\frac{2\pi\sqrt{(\rho)^2+(h/2\pi)^2}}{\Omega\sqrt{(\rho)^2+(h/2\pi)^2}} = \frac{2\pi}{\Omega},
\end{equation}
where $\Omega$ is the rotation between monomers in relation to the helical axis. Therefore, the length of the helical axis becomes $\dfrac{Nh}{m_t}$, from which the aspect ratio $\Lambda_{helix}$ for the entire polymer is given by:
\begin{equation}
    \Lambda_{helix} = \frac{Nh}{2m_t\rho}
\label{eqn: Helical Rod Aspect Ratio}
\end{equation}

To the authors' knowledge, equation~\ref{eqn: Helical Rod Aspect Ratio}  has not yet been reported in this continuous form. Yet a discrete version was offered by the multiscale density-functional theory introduced by Xu \textit{et al.}\cite{Xu2009} In this model, isotropic-to-nematic phase transitions of helical polymers on a hydrophobic surface were achieved \textit{via} a $\Lambda_{helix} \geq2$, demonstrating the dependence of liquid-crystalline behavior on this parameter. Additionally, in recent work, it was shown that for $\Lambda_{helix} \geq 7$ and large enough $K_\theta$ and $K_\phi$, lamellae-forming block copolymers also underwent a nematic transition.\cite{SpringsInStripes} \\

\begin{figure}[ht]
    \centering
    \includegraphics[width=1.0\linewidth]{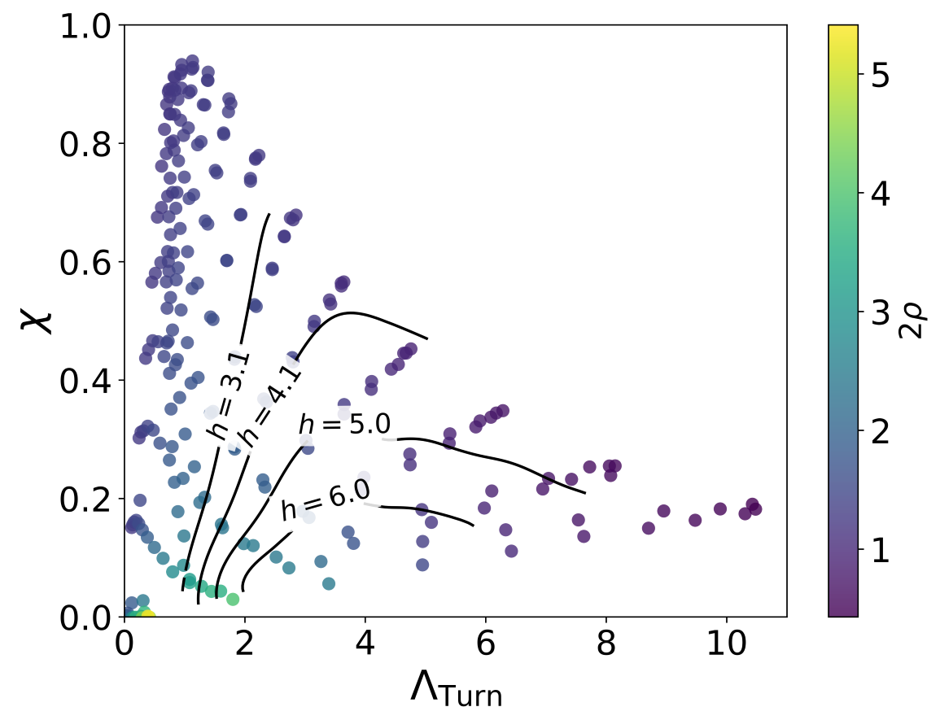}
    \caption{The tradeoff between chirality $\chi$ and helical aspect ratio $\Lambda_{turn}$ as a function of pitch $h$ and diameter $2\rho$.}
    \label{fig: Comparing Chirality and Aspect Ratio}
\end{figure}

Since $N$ also affects the total chain chirality ($\chi_0N$), control over orientational ordering requires careful consideration when one is also interested in simultaneously managing liquid-crystalline ordering and chirality. Specifically, $\Lambda_{turn}$ depends on the same variables as $\chi_0$ and cannot be independently tuned. Rather, we examine their correlation in Figure \ref{fig: Comparing Chirality and Aspect Ratio}, where $\Lambda_{turn}$ is compared to $\chi$; the contour lines are drawn based on pitch and the colors correspond to helical radius. While a decrease in diameter generally increases both $\chi$ and $\Lambda_{turn}$, the overall magnitude of this change is strongly dictated by the pitch. Smaller pitches lead to a greater chirality characteristic while simultaneously reducing $\Lambda_{turn}$. This trade-off cannot be simply compensated for by an appropriate design parameter, as increasing the degree of polymerization $N$ directly impacts the chiral response of the material.\cite{Okamoto2015, Jose2023}.

\section{Mapping real polymers to Kremer-Grest Helical Chain}
Given the ability of the Kremer–Grest helical chain to quantify a diverse range of chiral and helical conformations, it is necessary to be able to map these parameters to specific experimental systems. Here, we present a coarse-graining procedure that enables direct comparison between experimental (and, in some cases, \textit{ab initio} or all-atom simulations) polymer systems and the Kremer–Grest helical chain. To demonstrate the generality of the framework, we examine a diverse set of polymer classes, including poly(L-lactic acid) (PLLA), the $\alpha$-crystalline form of isotactic polypropylene (iPP), a poly(alkoxyallene), an ester-substituted polyfuran, and a polypeptoid, spanning a broad range of chemical architectures and helical propensities.\cite{Ho2009,Chavez2025, Zhou2021, Varni2019, Armand1998}

\subsection{Geometry-based Parameters for Kremer-Grest Helical Chains}
\begin{table*}
\centering
    \begin{tabular}{l c c c c c c c c c c c c }
    \toprule
    Helix 
    & Density ($g/cm^3$) 
    & $M_r$  ($g/mol$)
    & $l_0$ ($nm$)
    & $h_{exp}$ ($nm$) 
    & $\rho_{exp}$ ($nm$)
    & $h_{sim}$ ($\sigma$)
    & $\rho_{sim}$ ($\sigma$) 
    & $\kappa_0$
    & $\tau_0$
    & $\theta_0$
    & $\phi_0$
    & $\chi_0$\\
    \midrule

    PLLA 
    & 1.24 
    & 72 
    & 0.459 
    & 0.866 
    & 0.533 
    & 1.889 
    & 1.162 
    & 0.806 
    & 0.209 
    & $134^\circ$ 
    & $10^\circ$ 
    & 0.1486 \\
    
    iPP 
    & 0.946 
    & 42 
    & 0.419 
    & 0.660 
    & 0.386 
    & 1.574 
    & 0.921 
    & 1.011 
    & 0.275 
    & $121^\circ$ 
    & $17^\circ$ 
    & 0.3074 \\
    
    Polyallene 
    & 1.006 
    & 415 
    & 0.882 
    & 2.520 
    & 1.085 
    & 2.859 
    & 1.231 
    & 0.715 
    & 0.264 
    & $139^\circ$ 
    & $11^\circ$ 
    & 0.1479 \\
    
    P3HEF 
    & 1.12 
    & 196 
    & 0.662 
    & 0.340 
    & 2.270 
    & 0.513 
    & 3.427 
    & 0.292 
    & 0.007 
    & $164^\circ$ 
    & $0^\circ$ 
    & 0.0006 \\
    
    Polypeptoid 
    & 1.18 
    & 148 
    & 0.593 
    & 0.600 
    & 0.563 
    & 1.012 
    & 0.950 
    & 1.023 
    & 0.173
    & $121^\circ$ 
    & $11^\circ$ 
    & 0.1989 \\
        
    \bottomrule
    \end{tabular}
    \caption{The experimental values of Density ($g/cm^3$), repeat-unit molecular weight ($M_r$ $g/mol$) and monomer diameter ($l_0$ $nm$). These are used to create a relative dimensionless length scale ($\sigma$) between all polymers with PLLA as the reference. We further map the experimental pitch ($h_{exp}$ ($nm$)) and radius ($\rho$ ($nm$))) to their dimensionless counterparts, $h_{sim}$ and $\rho_{sim}$ respectively. Finally, we compute the equilibirum curvature ($\kappa_0$),  torsion ($\tau_0$) and associated bond angle ($\theta_0$), dihedral angle($\phi_0$) and chirality characteristic ($\chi_0$).}
  \label{tbl:setpoints}
\end{table*}
We start our coarse-graining by assuming that each monomer is represented as an isotropic sphere with a bead diameter $l_0$ (given in $nm$), that maps to the dimensionless length scale $\sigma$. To derive $l_0$, we compute the ratio of the molecular weight of one repeat unit ($M_r$) to the polymer density, and divide by Avogadro's number. Upon conversion to $nm^3$, the cube root is computed to yield the desired bead diameter in real units.\\

Next, we scale the experimental or \textit{in silico} (quantum mechanical or all atom simulations) computed pitch ($h_{exp}$) and radius ($\rho_{exp}$) by this length scale to yield a dimensionless pitch ($h_{sim}=h_{exp}/l_0$) and radius ($\rho_{sim}=\rho_{exp}/l_0$). While the literature often reports $h_{exp}$ and $m_t$, it typically fails to explicitly report the corresponding radius. In this scenario, one can infer the radius using the procedure outlined in Appendix E. We summarize the end result here as:

\begin{equation}
\rho = \sqrt{\frac{(l_0R_0)^2 - (h/m_t)^2}{2(1-\cos\Omega)}},
\label{eqn: rho for experiments}
\end{equation} 
This allows us to make use of equation \ref{eqn: parametric kappa and tau} to compute the equilibrium $\kappa_0$ and $\tau_0$ relevant to the Kremer-Grest helical chain. Finally, rearrangement of equations \ref{eqn: kappa} and \ref{eqn: tau} gives the bond and dihedral angle setpoints required for simulation:
\begin{equation}
    \label{eqn: backmapping kappa}
    \theta_0=cos^{-1}(\frac{\kappa_0^2R_0^2}{2}-1),
\end{equation}
\begin{equation}
    \label{eqn: backmapping tau}
    \phi_0=sin^{-1}(\frac{\tau_0\kappa_0^2R_0^3}{sin(\theta_0)}),
\end{equation}
This procedure was conducted for each of the aforementioned polymers, and the results are reported in Table \ref{tbl:setpoints}.\\

For the diverse set of chemistries, the curvatures span $\kappa_0\in[0.292, 0.985]$, with the conjugated P3HEF having the lowest $\kappa_0$. A conjugated polymer will typically have a much smaller curvature than its non-conjugated counterparts due to the energetically favorable $\pi$-orbital overlap associated with its $sp_2$ hybridization. Conversely, the polypeptoid has the largest curvature, followed closely by iPP. The derived torsion setpoints fall within the range of $\tau_0\in[0.007, 0.264]$. Again, P3HEF represents the minimum value in this set due to its conjugated backbone promoting a planar structure, while the polyallene takes on the largest $\tau$. These setpoints are particularly promising because they correspond to the regime of low error from our theoretical model. Specifically, low error in $\kappa$ and $\tau$ was observed on the bonds of $\theta_0\in[120^\circ, 140^\circ]$ with $\phi_0\in[0^\circ,15^\circ]$.\\

Finally, we show the computed $\chi_0$ and observe that all the polymers have relatively similar and modest values of chirality except for P3HEF, which is essentially achiral due to its planar structure. Interestingly, iPP, one of the structurally simplest polymer chains, contains the largest value. Although crystallization is required for this helical conformation to emerge, this observation demonstrates that chirality can arise from relatively simple molecular interactions. To illustrate the mapping procedure, we perform simulations at the prescribed equilibrium bond and dihedral angles. The force constants governing these setpoints are fixed at $2K_\theta=K_\phi=10k_BT$. \\

\begin{figure}[ht]
    \centering
    \includegraphics[width=0.8\linewidth]{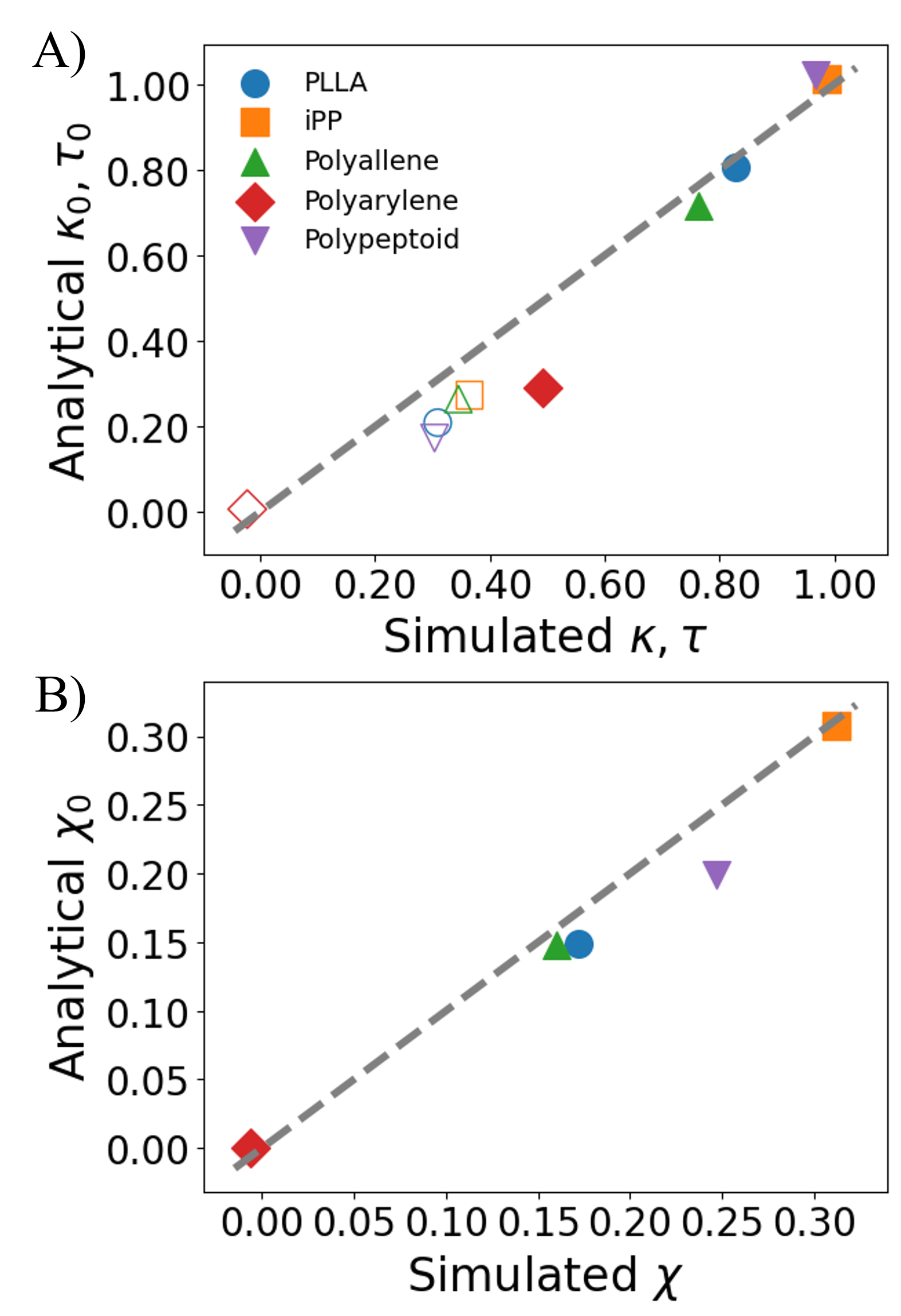}
    \caption{(A) Comparison between simulated $\kappa$ (filled markers) and $\tau$ (open markers), and the experimental values for all five polymers. (B) Comparison between simulated $\chi$ and the experimental values for all five polymers.}
    \label{fig: Comparing Experiment to Simulation}
\end{figure}

Figure~\ref{fig: Comparing Experiment to Simulation}A quantifies the deviation between experimentally measured and simulated $\kappa$ and $\tau$, revealing strong agreement between the Kremer-Grest helical model and the experimentally derived values. Notably, the polyarylene exhibited the largest difference from its experimentally derived $\kappa_0$. This is expected, as our simulation only accounts for repulsive pairwise interactions while the polyarylene conformation promotes $\pi$-stacking, which is an attractive force. This results in a $h_{sim}$ of $0.513\sigma$, which is well below the $2^{1/6}\sigma$ cutoff of the WCA potential, ultimately extending the polymer to alleviate these tightly packed helical loops, in turn increasing the curvature. On the other hand, the torsion of the polyarylene remains relatively unperturbed, as the large bond angle between subsequent monomers prevents any 1-4 pairwise interaction regardless of the planar conformation. For all other polymers, we observe a simulated $\tau$ that is systematically larger than the expected $\tau_0$, albeit to a small degree. \\

Furthermore, Figure~\ref{fig: Comparing Experiment to Simulation}B shows that the simulated chirality characteristic $\chi$ is in fairly good agreement with the experimentally derived value across a diverse array of polymer classes, which demonstrates the robustness and efficacy of the proposed framework. Accordingly, we envision that this model will be used to rationally engineer polymeric systems with controlled chirality, enabling access to hierarchical chiral organization and facilitating advances in chiroptical technologies.\\

We have presented a simple geometry-based coarse-graining procedure that can faithfully capture the helical geometries of experimental systems, but we note that improvements for specific chemistries can be made by incorporating additional features. Attaching side chain beads, including specific attractive interactions, increasing the stiffness of the backbone, and regulating bond distances and bead sizes can all be leveraged for better agreement with experiment. Deeper knowledge of experimental measures of chirality, such as circular dichroism, can also be used to interrogate the predictions of chirality characteristic ($\chi_0$).

\section{Conclusion} 
In this work, a general framework is established for the quantitative measurement of the geometry of helical polymers and the associated chirality characteristic. Specifically, disparate approaches to thinking about helical polymers such as the helical wormlike chain model in terms of curvature and torsion $(\kappa_0,\tau_0)$, a geometric description of a helix in terms of helical pitch and radius $(h_0,\rho_0)$, and the Kremer-Grest helical model in terms of the angular and dihedral setpoints $(\theta_0,\phi_0)$ are examined and a prescription for the one-to-one mapping between viewpoints is established. The mapping enables the selection of appropriate models to study the physics of helical polymers. In particular, we have introduced a transferable coarse-grained framework that enables quantitative mapping between the geometry of experimentally realized helical polymers and the Kremer–Grest helical chain model. By systematically relating experimentally accessible geometric parameters to effective bond and dihedral setpoints, the model captures the magnitude and handedness of polymer chirality while explicitly accounting for excluded volume effects and thermal fluctuations. Across a diverse range of polymer classes—including PLLA, isotactic polypropylene, polyallenes, polyfurans, and polypeptoids—the simulated curvature and torsion show strong agreement with experimental and \textit{in silico} reference values, highlighting both the robustness and generality of the approach. \\ 

We further make a connection between the geometry of the helix and its possible impact on liquid-crystalline self-assembly behavior through the aspect ratio per turn, and on chiral properties through the chirality characteristic. We envision that the framework will be used for the inverse design of polymeric systems with targeted chiral and mesoscale properties. Desired macroscopic behaviors can be specified \textit{a priori} and translated into molecular-scale geometric constraints, which enable efficient exploration of chemical space to identify polymer architectures that yield prescribed chiral strength, as well as selective stabilization of liquid-crystalline phases. More broadly, the ability to decouple geometric design from chemical complexity establishes a versatile platform for engineering chirality-driven functionality, including tunable optical activity, enantioselective transport, and programmable self-assembly in soft matter systems.

\begin{acknowledgments}
The work performed for this manuscript was funded and supported by the National Science Foundation DMR-2144511 and the Kate Gleason College Fund. The authors acknowledge Research Computing at RIT for computational resources and support in compiling and managing software libraries utilized in this work.
\end{acknowledgments}

\section*{Data Availability Statement}
The data that supports the findings of this manuscript are available upon request to the corresponding authors.

\appendix

\section{Derivation of $(\kappa_0,\tau_0) \leftrightarrow (\theta_0,\phi_0)$}

We begin by explicitly expanding the curvature in terms of the discrete tangent vectors:
\begin{equation}
\begin{aligned}
\kappa 
&= 
\left| \frac{d\mathbf{t}_i}{R_0} \right| 
= 
\frac{1}{R_0} 
\Bigg[ 
(\mathbf{t}_{i,x}-\mathbf{t}_{i-1,x})^2
+ (\mathbf{t}_{i,y}-\mathbf{t}_{i-1,y})^2 \\
&\quad 
+ (\mathbf{t}_{i,z}-\mathbf{t}_{i-1,z})^2
\Bigg]^{1/2},
\label{eqn: A1_split}
\end{aligned}
\end{equation}
and, upon expanding the squares, we obtain
\begin{equation}
\begin{aligned}
\kappa
&= \frac{1}{R_0}
\Bigg[
\mathbf{t}_{i,x}^2 + \mathbf{t}_{i-1,x}^2
+ \mathbf{t}_{i,y}^2 + \mathbf{t}_{i-1,y}^2
+ \mathbf{t}_{i,z}^2 + \mathbf{t}_{i-1,z}^2 \\
&\quad
- 2 \left(
\mathbf{t}_{i,x} \mathbf{t}_{i-1,x}
+ \mathbf{t}_{i,y} \mathbf{t}_{i-1,y}
+ \mathbf{t}_{i,z} \mathbf{t}_{i-1,z}
\right)
\Bigg]^{1/2}.
\label{eqn: A2}
\end{aligned}
\end{equation}
Next, by recognizing the dot product between successive unit tangent vectors as the cosine of the angle $\theta$ between them, we have
\begin{equation}
\cos(\theta) = \mathbf{t}_i \cdot \mathbf{t}_{i-1} 
= \mathbf{t}_{i,x} \mathbf{t}_{i-1,x}
+ \mathbf{t}_{i,y} \mathbf{t}_{i-1,y}
+ \mathbf{t}_{i,z} \mathbf{t}_{i-1,z}.
\end{equation}
Substituting this expression into equation~\eqref{eqn: A2} yields
\begin{equation}
\kappa
= \frac{1}{R_0}
\sqrt{
\mathbf{t}_{i,x}^2 + \mathbf{t}_{i-1,x}^2
+ \mathbf{t}_{i,y}^2 + \mathbf{t}_{i-1,y}^2
+ \mathbf{t}_{i,z}^2 + \mathbf{t}_{i-1,z}^2
- 2 \cos(\theta)
}.
\label{eqn: A5}
\end{equation}
Finally, noting that $\mathbf{t}_i^2 = \mathbf{t}_i \cdot \mathbf{t}_i = \mathbf{t}_{i,x}^2 + \mathbf{t}_{i,y}^2 + \mathbf{t}_{i,z}^2$, we can simplify equation~\eqref{eqn: A5} to
\begin{equation}
\kappa
= \frac{1}{R_0} \sqrt{ \mathbf{t}_i^2 + \mathbf{t}_{i-1}^2 - 2 \cos(\theta) }.
\end{equation}
For unit tangent vectors ($\mathbf{t}_i^2 = \mathbf{t}_{i-1}^2 = 1$), this reduces to the familiar form
\begin{equation}
\kappa_0 = \frac{\sqrt{2 \left[ 1 - \cos(\theta_0) \right]}}{R_0}.
\end{equation}
In our model, the sign of the trigonometric function is taken to be positive to be consistent with the method used in LAMMPS for computing bond angles.

For the derivation of $\tau_0$, we start with the definition of torsion:
\begin{equation}
\tau = -\frac{\mathbf{b}_i - \mathbf{b}_{i-1}}{R_0} \cdot \mathbf{n}_i.
\end{equation}
Substituting the definition of the unit normal, $\mathbf{n}_i = \frac{\mathbf{t}_i - \mathbf{t}_{i-1}}{|\mathbf{t}_i - \mathbf{t}_{i-1}|}$:
\begin{equation}
\tau = -\frac{\mathbf{b}_i - \mathbf{b}_{i-1}}{R_0} 
\cdot \frac{\mathbf{t}_i - \mathbf{t}_{i-1}}{|\mathbf{t}_i - \mathbf{t}_{i-1}|}.
\end{equation}
Using $\kappa = \frac{|\mathbf{t}_i - \mathbf{t}_{i-1}|}{R_0}$:
\begin{equation}
\tau = -\frac{\mathbf{b}_i - \mathbf{b}_{i-1}}{R_0} 
\cdot \frac{\mathbf{t}_i - \mathbf{t}_{i-1}}{\kappa \, R_0}.
\end{equation}
Substituting the definition of the binormal:
\begin{equation}
\begin{aligned}
\tau &= -\frac{1}{R_0^3 \, \kappa^2} \Big[
(\mathbf{t}_i \times (\mathbf{t}_i - \mathbf{t}_{i-1})) \\
&\quad - (\mathbf{t}_{i-1} \times (\mathbf{t}_{i-1} - \mathbf{t}_{i-2}))
\Big] \cdot (\mathbf{t}_i - \mathbf{t}_{i-1}).
\end{aligned}
\end{equation}
Simplifying cross products:
\begin{equation}
\begin{aligned}
\tau &= -\frac{1}{R_0^3 \, \kappa^2} 
\left[ -(\mathbf{t}_i \times \mathbf{t}_{i-1}) + (\mathbf{t}_{i-1} \times \mathbf{t}_{i-2}) \right] 
\cdot (\mathbf{t}_i - \mathbf{t}_{i-1}) \\
&= -\frac{1}{R_0^3 \, \kappa^2} (\mathbf{t}_i - \mathbf{t}_{i-1}) 
\cdot (\mathbf{t}_{i-1} \times \mathbf{t}_{i-2}).
\end{aligned}
\end{equation}
Expanding the dot product:
\begin{equation}
\tau = -\frac{1}{R_0^3 \, \kappa^2} 
\left[ \mathbf{t}_i \cdot (\mathbf{t}_{i-1} \times \mathbf{t}_{i-2}) 
- \mathbf{t}_{i-1} \cdot (\mathbf{t}_{i-1} \times \mathbf{t}_{i-2}) \right].
\end{equation}
The second term vanishes:
\begin{equation}
\tau = -\frac{1}{R_0^3 \, \kappa^2} \, \mathbf{t}_i \cdot (\mathbf{t}_{i-1} \times \mathbf{t}_{i-2}).
\label{eqn:tauDerivation}
\end{equation}
Expressing scalar triple product in terms of cosine:
\begin{equation}
\tau = -\frac{1}{R_0^3 \, \kappa^2} \, 
|\mathbf{t}_i| \, |\mathbf{t}_{i-1} \times \mathbf{t}_{i-2}| \, \cos(\psi).
\end{equation}
By writing the cross product in terms of sine and given that the magnitude of each unit vector is unity, this reduces to:
\begin{equation}
\tau = -\frac{1}{R_0^3 \, \kappa^2} \, \sin(\theta) \, \cos(\psi).
\end{equation}
Projecting onto the plane orthogonal to $\mathbf{t}_{i-1}$ we observe that $\psi = \phi - \pi/2$ which allows us to write:
\begin{equation}
\tau = -\frac{1}{R_0^3 \, \kappa^2} \, \sin(\theta) \, \cos(\phi - \pi/2)
= -\frac{1}{R_0^3 \, \kappa^2} \, \sin(\theta) \, \sin(\phi).
\end{equation}
Finally, solving for $\phi$:
\begin{equation}
\phi = \arcsin \left( \frac{\tau \, \kappa^2 \, R_0^3}{\sin(\theta)} \right).
\end{equation}

\section{Analytical 1-3 Pairwise Distance}
The distance between particles with relative indices of 1 and 3 in terms of $\theta$ requires use of the law of cosines. Specifically, if $|\mathbf{r_{13}}|=|\mathbf{r}_{i}-\mathbf{r}_{i-2}|$ then we can compute:

\begin{equation}
    |\mathbf{r}_{13}|=\sqrt{\mathbf{r}_{12}^2 + \mathbf{r}_{23}^2-2\mathbf{r}_{12}\mathbf{r}_{23}cos(\theta)},
    \label{eqn: SAS}
\end{equation}
In the analytical model, $|\mathbf{r}_{12}|=|\mathbf{r}_{23}|=R_0$, which upon substitution and through algebraic manipulation yields:

\begin{equation}
    |\mathbf{r}_{13}|=\sqrt{2R_0^2[1-cos(\theta)]},
    \label{eqn: r13}
\end{equation}

The Kremer-Grest helical chain uses a pairwise cutoff length of $r_c=2^{1/6}\sigma$. Substituting this into equation \ref{eqn: r13} and solving for $\theta$ gives the maximum bond angle ($\theta_c$) in which these particles will interact:

\begin{equation}
    \label{eqn: cos_theta_r13}
    \theta _c= cos^{-1}(1-r_c^2/2R_0)
\end{equation}

\section{Analytical 1-4 Pairwise Distance}
Computing the interactions between monomers with relative indices of 1 and 4 as a function of $\theta$ and $\phi$ requires defining the end to end vector:

\begin{equation}
    \mathbf{r}_{14} = \mathbf{r}_{12} + \mathbf{r}_{23} +  \mathbf{r}_{34}.
\end{equation}
Taking the magnitude of both sides of this equation, and substituting the cosine form of the dot products yields:

\begin{equation}
\begin{aligned}
|\mathbf{r}_{14}|^2 =\;& 
|\mathbf{r}_{12}|^2 
+ |\mathbf{r}_{23}|^2
+ |\mathbf{r}_{34}|^2 \\
&+ 2|\mathbf{r}_{12}||\mathbf{r}_{23}|\cos(\theta_{123}) \\
&+ 2|\mathbf{r}_{23}||\mathbf{r}_{34}|\cos(\theta_{234}) \\
&+ 2\,\mathbf{r}_{12}\cdot\mathbf{r}_{34},
\label{eqn: D2}
\end{aligned}
\end{equation}
where the final dot product was left in place as it requires special consideration. Since $\mathbf{r}_{12}$ and $\mathbf{r}_{34}$ share the intermediate bond $\mathbf{r}_{23}$, it is convenient to define each vector in terms of its parallel and orthogonal components:

\begin{equation}
    \mathbf{r}_{ij}=\mathbf{r}_{ij}^{\parallel}+\mathbf{r}_{ij}^{\perp}.
\end{equation}
This allows us to write the dot product as:

\begin{equation}
    \mathbf{r}_{12}\cdot \mathbf{r}_{34}=(\mathbf{r}_{12}^{\parallel}+\mathbf{r}_{12}^{\perp}) \cdot (\mathbf{r}_{34}^{\parallel}+\mathbf{r}_{34}^{\perp}).
    \label{eqn: D4}
\end{equation}
Given the shared bond $\mathbf{r}_{23}$, all parallel components are along $\mathbf{r}_{23}$ and all perpendicular components are orthogonal to $\mathbf{r}_{23}$. Therefore, equation \ref{eqn: D4} becomes:

\begin{equation}
    \mathbf{r}_{12}\cdot \mathbf{r}_{34}=\mathbf{r}_{12}^{\parallel}\cdot\mathbf{r}_{34}^{\parallel} + \mathbf{r}_{12}^{\perp}\cdot\mathbf{r}_{34}^{\perp}.
    \label{eqn: D5}
\end{equation}
We then define each component in terms of their projection along $\mathbf{r}_{23}$:

\begin{equation}
    \mathbf{r}_{12}^\parallel=\mathbf{r}_{12}cos(\theta_{123}),
\end{equation}
\begin{equation}
    \mathbf{r}_{12}^\perp=\mathbf{r}_{12}sin(\theta_{123}),
\end{equation}
\begin{equation}
    \mathbf{r}_{34}^\parallel=\mathbf{r}_{34}cos(\theta_{234}),
\end{equation}
\begin{equation}
    \mathbf{r}_{34}^\perp=\mathbf{r}_{34}sin(\theta_{234}).
\end{equation}
The perpendicular components lie in a plane orthogonal to $\mathbf{r}_{23}$ where the angle between them is the dihedral angle $\phi$. Substitution into equation \ref{eqn: D5} yields:

\begin{equation}
\begin{aligned}
\mathbf{r}_{12}\cdot \mathbf{r}_{34}
&=
|\mathbf{r}_{12}||\mathbf{r}_{34}|
\cos(\theta_{123})\cos(\theta_{234}) \\
&\quad+
|\mathbf{r}_{12}||\mathbf{r}_{34}|
\sin(\theta_{123})\sin(\theta_{234})\cos(\phi).
\label{eqn: D10}
\end{aligned}
\end{equation}
Due to the theoretical nature of the model, we can define all bond distances as $|\mathbf{r}_{12}|=|\mathbf{r}_{23}|=|\mathbf{r}_{34}|=R_0$ and all bond angles as $\theta_{123}=\theta_{234}=\theta$. Then, substitution of equation \ref{eqn: D10} into \ref{eqn: D2} gives:

\begin{equation}
|\mathbf{r}_{14}|
=
\sqrt{
R_0^2
\left[
3
+ 4\cos(\theta)
+ 2\cos^2(\theta)
+ 2\sin^2(\theta)\cos(\phi)
\right]}.
\end{equation}

\section{Estimation of Helical Radius}
We wish to compute the radius $\rho$ of the helix by assuming that the first bead ($\mathbf{r}_1$) lies in the $xy$-plane some distance $\rho$ from the origin. The next bead ($\mathbf{r}_2$) is located a distance $l_0R_0$ away with the per monomer helical rise in the $z$-direction being given as $h/m_t$. Projection of $\mathbf{r}_2$ onto a circle transcribed in the $xy$-plane gives a right triangle with vertices at each bead position and this newly projected point $\mathbf{\zeta}$. Using the Pythagorean Theorem, we can solve for the distance between $\mathbf{r}_1$ and $\zeta$ as:

\begin{equation}
    \zeta=\sqrt{(l_0R_0)^2-(h/m_t)^2}.
\end{equation}
This yields a triangle in the $xy$-plane with vertices at the origin, $\mathbf{r}_1$ and $\mathbf{\zeta}$, where the distance between $\mathbf{\zeta}$ and the origin is also $\rho$. Recall that $\Omega=2\pi/m_t$ and defines the angle of rotation between monomer. This allows us to again leverage the Law of Cosines to solve for $\rho$:

\begin{equation}
    \mathbf{\zeta}^2=2\rho^2[1-cos(\Omega)].
\end{equation}
Substituting $\mathbf{\zeta}$:
\begin{equation}
    \sqrt{(l_0R_0)^2-(h/m_t)^2}^2=2\rho^2[1-cos(\Omega)],
\end{equation}
followed by rearrangement and takings the square leads to the final estimation for $\rho$:

\begin{equation}
    \rho = \sqrt{\frac{(l_0R_0)^2-(h/m_t)^2}{2[1-cos(\Omega)}}
\end{equation}
\nocite{*}
\bibliography{citations}

\end{document}